\documentclass[english,aps,preprint]{revtex4}
\usepackage[T1]{fontenc}
\usepackage[utf8]{inputenc}
\usepackage{amsmath}
\usepackage{amssymb}

\makeatletter

\usepackage{bm}

\makeatother

\usepackage{babel}

\begin{document}

\title{Correction to the Moli\`ere's formula for multiple scattering.}

\author{R.N. Lee}

\email[Email:]{R.N.Lee@inp.nsk.su}

\author{A.I. Milstein}

\email[Email:]{A.I.Milstein@inp.nsk.su}

\affiliation{Budker Institute of Nuclear Physics \\
 and\\
 Novosibirsk State University,\\
 630090 Novosibirsk, Russia}

\date{\today}
\begin{abstract}
The quasiclassical correction to the Moli\`ere's formula for multiple scattering is derived. The consideration is based
on the scattering amplitude, obtained with the first quasiclassical correction taken into account for arbitrary
localized but not spherically symmetric potential. Unlike the leading term, the correction to the Moli\`ere's formula
contains the target density $n$ and thickness $L$ not only in the combination $nL$ (areal density). Therefore, this
correction can be reffered to as the bulk density correction. It turns out that the bulk density correction is small
even for high density. This result explains the wide region of applicability of the Moli\`ere's formula.
\end{abstract}

\pacs{d}

\maketitle

\section{Introduction}

Multiple scattering of high-energy particles in matter is a process which plays an important role in the experimental
physics. The basis of the theoretical description of this process goes back to Refs.
\cite{Goudsmit1940a,Goudsmit1940,Moli`ere1948,Bethe1953}. Further development of the theory of multiple scattering was
performed in numerous publications, see, e.g., Refs. \cite{Fernandez-Varea1993,Negreanu2005} and references therein.
Detailed experimental investigation of multiple scattering has also been performed, see
Refs.\cite{Hanson1951,Ross2008}.

The celebrated Moli\`ere's formula describes the angular distribution $\frac{dW}{d\Omega}$ for small-angle scattering.
It was shown by Bethe in Ref. \cite{Bethe1953} that the most simple way to derive this formula is to use the transport
equation. As a consequence of this equation, the angular distribution $\frac{dW}{d\Omega}$ depends on the thickness $L$
and the density $n$ only in the combination $nL$, which is the areal density of a target. One can expect that the
applicability of the Moli\`ere's formula is restricted by the low density. However, the experimental results obtained
for small scattering angles show that the deviations from the Moli\`ere's formula are small for all data available. In
the present paper we explain such surprising behavior calculating the leading bulk-density correction to the
Moli\`ere's formula.

We start with the expression for the small-angle scattering amplitude. This expression has been obtained in Ref.
\cite{Lee2000b} in the quasiclassical approximation with the first correction taken into account. The applicability of
this approximation is provided by small scattering angles and high energy $\varepsilon$ of the particle,
$\varepsilon\gg m$ ($m$ is the particle mass, the system of units with $\hbar=c=1$ is used). This amplitude has been
obtained for arbitrary localized potential without requirement of its spherical symmetry. As known, the quasiclassical
wave function has a much wider region of applicability than the eikonal wave function. However, as it was shown in Ref.
\cite{Lee2000b}, the scattering amplitude obtained with the use of the quasiclassical wave function coincides with that
obtained in the eikonal approximation, see also Ref. \cite{Akhiezer1975}. Using the quasiclassical scattering amplitude
with the first correction taken into account, we calculate the corresponding cross section and average it over the
positions of atoms in the target. Dividing this cross section by the area of the target, we arrive at the angular
distribution $\frac{dW}{d\Omega}$. The leading term of this distribution coincides with the Moli\`ere's formula. The
correction depends not only on the areal density $nL$ of the target but also on the bulk density $n$ alone. We discuss
the magnitude of the correction for different target parameters and scattering angles.

\section{Differential probability}

Let us direct the $z$ axis along the initial momentum $\mathbf{p}$ of the particle so that
$\mathbf{r}=z\,\mathbf{p}/p+\boldsymbol{\xi}$. The small-angle high-energy scattering amplitude in the localized
potential $V\left(z,\boldsymbol{\xi}\right)$ has the form \cite{Lee2000b}
\begin{eqnarray}
f & = & -\frac{i\varepsilon}{2\pi}\int d^{2}\xi\boldsymbol{\,}e^{-i\mathbf{q}\cdot\boldsymbol{\xi}}\Biggl\{
e^{-i\mathcal{K}\left(\boldsymbol{\xi}\right)}-1+e^{-i\mathcal{K}\left(\boldsymbol{\xi}\right)}
\Biggl[\frac{1}{2\varepsilon}\int\limits _{-\infty}^{\infty} dx\, x\triangle_{\boldsymbol{\xi}}V\left(x,\boldsymbol{\xi}\right)\nonumber \\
 &  & -\frac{i}{\varepsilon}\int\limits _{-\infty}^{\infty}dx\int\limits _{-\infty}^{x}dy\, y\,
 \left(\nabla_{\boldsymbol{\xi}}V\left(x,\boldsymbol{\xi}\right)\right)\cdot\left(\nabla_{\boldsymbol{\xi}}
 V\left(y,\boldsymbol{\xi}\right)\right)\Biggr]\Biggr\},\label{eq:ampl}
 \end{eqnarray}
where $\mathbf{q}=\mathbf{p}'-\mathbf{p},$ $\mathbf{p}'$ is the
final momentum, $\mathcal{K}\left(\boldsymbol{\xi}\right)=\int_{-\infty}^{\infty}dxV\left(x,\boldsymbol{\xi}\right)$,
$\nabla_{\boldsymbol{\xi}}=\partial/\partial\boldsymbol{\xi}$, and
$\triangle_{\boldsymbol{\xi}}=\nabla_{\boldsymbol{\xi}}^{2}$. The
second term in braces in Eq. (\ref{eq:ampl}) corresponds to the correction.
For $\mathbf{q}\neq0$, the unity in the leading term can be omitted.
The differential cross section, corresponding to the amplitude $f$
and having the same accuracy as Eq. (\ref{eq:ampl}), reads\begin{gather}
\frac{d\sigma}{d\Omega}=\frac{\varepsilon^{2}}{2\pi^{2}}\mathrm{Re}\int d\boldsymbol{\xi}_{1}\, d\boldsymbol{\xi}_{2}e^{-i\mathbf{q}\cdot\left(\boldsymbol{\xi}_{1}-\boldsymbol{\xi}_{2}\right)}e^{-i\left[\mathcal{K}\left(\boldsymbol{\xi}_{1}\right)-\mathcal{K}\left(\boldsymbol{\xi}_{2}\right)\right]}\nonumber \\
\times\left\{ 1+\frac{1}{2\varepsilon}\int\limits _{-\infty}^{\infty} dx\,
x\triangle_{\boldsymbol{\xi}_{1}}V\left(x,\boldsymbol{\xi}_{1}\right)-\frac{i}{\varepsilon}\int\limits
_{-\infty}^{\infty}dx\int\limits
_{-\infty}^{x}dyy\,\left(\nabla_{\boldsymbol{\xi}_{1}}V\left(x,\boldsymbol{\xi}_{1}\right)\right)\cdot\left(\nabla_{\boldsymbol{\xi}_{1}}V\left(y,\boldsymbol{\xi}_{1}\right)\right)\right\}
.\label{eq:-1}\end{gather}
 The total potential of atoms in the target has the form

\begin{equation}
V\left(\boldsymbol{r}\right)=\sum_{i}u\left(\boldsymbol{r}-\boldsymbol{r}_{i}\right),\label{eq:}\end{equation}
where $u\left(\mathbf{r}\right)$ is the potential of individual atom,
which we assume to be spherically symmetric. We perform the averaging
over the atomic positions using the prescription\begin{equation}
\left\langle f\right\rangle =\int\prod_{i}\frac{dx_{i}d\boldsymbol{\rho}_{i}}{LS}f,\label{eq:}\end{equation}
corresponding to the dilute gas approximation. As a result, we obtain
\begin{align}
\frac{dW}{d\Omega}=\left\langle \frac{d\sigma}{Sd\Omega}\right\rangle  & =\frac{\varepsilon^{2}}{2\pi^{2}}\mathrm{Re}\int d\boldsymbol{\rho}e^{-i\mathbf{q}\cdot\boldsymbol{\rho}}\Biggl\{ F_{1}^{N}-\frac{iN}{\varepsilon}F_{1}^{N-1}F_{2}-\frac{iN\left(N-1\right)}{\varepsilon}F_{1}^{N-2}F_{3}\Biggr\},\label{eq:W1}\end{align}
where\begin{eqnarray}
F_{1} & = & \int\frac{d\boldsymbol{\rho}_{1}}{S}\exp\left\{ -i\left[\chi\left(\boldsymbol{\rho}_{1}\right)-\chi\left(\boldsymbol{\rho}_{1}-\boldsymbol{\rho}\right)\right]\right\} ,\nonumber \\
F_{2} & = & \frac{L}{4}\int\frac{d\boldsymbol{\rho}_{1}}{S}\exp\left\{ -i\left[\chi\left(\boldsymbol{\rho}_{1}\right)-\chi\left(\boldsymbol{\rho}_{1}-\boldsymbol{\rho}\right)\right]\right\} \left[\left(\nabla_{\boldsymbol{\rho}_{1}}\chi\left(\boldsymbol{\rho}_{1}\right)\right)\cdot\left(\nabla_{\boldsymbol{\rho}_{1}}\chi\left(\boldsymbol{\rho}_{1}\right)\right)+i\triangle_{\boldsymbol{\rho}_{1}}\chi\left(\boldsymbol{\rho}_{1}\right)\right]\nonumber \\
 &  & +\int\limits _{-\infty}^{\infty}dx\int\limits _{-\infty}^{x}dy\, y\int\frac{d\boldsymbol{\rho}_{1}}{S}\exp\left\{ -i\left[\chi\left(\boldsymbol{\rho}_{1}\right)-\chi\left(\boldsymbol{\rho}_{1}-\boldsymbol{\rho}\right)\right]\right\} \left(\nabla_{\boldsymbol{\rho}_{1}}u\left(x,\boldsymbol{\rho}_{1}\right)\right)\cdot\left(\nabla_{\boldsymbol{\rho}_{1}}u\left(y,\boldsymbol{\rho}_{1}\right)\right),\nonumber \\
F_{3} & = & \int\limits _{-\infty}^{\infty}dx\int\limits _{-\infty}^{\infty}dy\iint\frac{d\boldsymbol{\rho}_{1}}{S}\frac{d\boldsymbol{\rho}_{2}}{S}\exp\left\{ -i\left[\chi\left(\boldsymbol{\rho}_{1}\right)-\chi\left(\boldsymbol{\rho}_{1}-\boldsymbol{\rho}\right)+\chi\left(\boldsymbol{\rho}_{2}\right)-\chi\left(\boldsymbol{\rho}_{2}-\boldsymbol{\rho}\right)\right]\right\} \nonumber \\
 &  & \times\left[\frac{L}{6}-\frac{\left(x-y\right)^{2}}{4L}\right]\left(\nabla_{\boldsymbol{\rho}_{1}}u\left(x,\boldsymbol{\rho}_{1}\right)\right)\cdot\left(\nabla_{\boldsymbol{\rho}_{2}}u\left(y,\boldsymbol{\rho}_{2}\right)\right).\label{eq:F1-4}\end{eqnarray}

Here $\chi\left(\boldsymbol{\rho}\right)=\int_{-\infty}^{\infty}dx\, u\left(x,\boldsymbol{\rho}\right)$,
so that $\mathcal{K}\left(\boldsymbol{\rho}\right)=\sum_{i}\chi\left(\boldsymbol{\rho}-\boldsymbol{\rho}_{i}\right)$.
Then we pass to the limit $N,\, S\to\infty$ and $N/S=nL=const$.
In this limit,\begin{eqnarray}
F_{1}^{N} & = & \left(1+\int\frac{d\boldsymbol{\rho}_{1}}{S}\left[\exp\left\{ -i\left[\chi\left(\boldsymbol{\rho}_{1}\right)-\chi\left(\boldsymbol{\rho}_{1}-\boldsymbol{\rho}\right)\right]\right\} -1\right]\right)^{N}\nonumber \\
 & \to & \exp\left[-nL\int d\boldsymbol{\rho}_{1}\left(1-e^{-i\left[\chi\left(\boldsymbol{\rho}_{1}\right)-\chi\left(\boldsymbol{\rho}_{1}-\boldsymbol{\rho}\right)\right]}\right)\right].\label{eq:F1N}\end{eqnarray}

Substituting Eqs. (\ref{eq:F1-4}) and (\ref{eq:F1N}) into Eq. (\ref{eq:W1}),
we finally obtain\begin{eqnarray}
\frac{dW}{d\Omega} & = & \frac{\varepsilon^{2}}{\left(2\pi\right)^{2}}\int d\boldsymbol{\rho}e^{-i\mathbf{q}\cdot\boldsymbol{\rho}}\exp\left\{ -nL\int d\boldsymbol{\rho}_{1}\left[1-\cos\left(\chi\left(\boldsymbol{\rho}_{1}\right)-\chi\left(\boldsymbol{\rho}_{1}-\boldsymbol{\rho}\right)\right)\right]\right\} \nonumber \\
 &  & \times\biggl\{1-\frac{nL}{\varepsilon}\int\limits _{-\infty}^{\infty}dx\int d\boldsymbol{\rho}_{1}\sin\left(\chi\left(\boldsymbol{\rho}_{1}\right)-\chi\left(\boldsymbol{\rho}_{1}-\boldsymbol{\rho}\right)\right)\boldsymbol{\rho}_{1}\cdot\nabla_{\boldsymbol{\rho}_{1}}u^{2}\left(x,\boldsymbol{\rho}_{1}\right)\nonumber \\
 &  & -\frac{n^{2}L}{\varepsilon}\int d\boldsymbol{\rho}_{1}\cos\left(\chi\left(\boldsymbol{\rho}_{1}\right)-\chi\left(\boldsymbol{\rho}_{1}-\boldsymbol{\rho}\right)\right)\boldsymbol{\rho}_{1}\chi\left(\boldsymbol{\rho}_{1}\right)\nonumber \\
 &  & \times\int d\boldsymbol{\rho}_{2}\,\sin\left(\chi\left(\boldsymbol{\rho}_{2}\right)-\chi\left(\boldsymbol{\rho}_{2}-\boldsymbol{\rho}\right)\right)\nabla_{\boldsymbol{\rho}_{2}}\chi\left(\boldsymbol{\rho}_{2}\right)\biggl\}.\label{eq:W2}\end{eqnarray}
 At the derivation of this formula we have used the integration by
parts. As a result, all terms proportional to $L$ in $F_{2}$ and $F_{3}$, Eq. (\ref{eq:F1-4}), vanished. It is
convenient to rewrite Eq. (\ref{eq:W2}) in another form. The differential in the momentum transfer $\mathbf{Q}$ cross
section $\frac{d\sigma}{d\mathbf{Q}}$ of high-energy scattering off one atom, calculated in the quasiclassical
approximation with the first correction taken into account, satisfies the relation \cite{Akhiezer1975,Lee2000b}
\begin{eqnarray}
\int d^{2}Q\left(1-e^{i\mathbf{Q}\cdot\boldsymbol{\rho}}\right)\frac{d\sigma}{d\mathbf{Q}} & = & \int d\boldsymbol{\rho}_{1}\left[1-\cos\left(\chi\left(\boldsymbol{\rho}_{1}\right)-\chi\left(\boldsymbol{\rho}_{1}-\boldsymbol{\rho}\right)\right)\right.\nonumber \\
 &  & \left.+\frac{1}{\varepsilon}\int\limits _{-\infty}^{\infty}dx\sin\left(\chi\left(\boldsymbol{\rho}_{1}\right)-\chi\left(\boldsymbol{\rho}_{1}-\boldsymbol{\rho}\right)\right)\boldsymbol{\rho}_{1}\cdot\nabla_{\boldsymbol{\rho}_{1}}u^{2}\left(x,\boldsymbol{\rho}_{1}\right)\right].\label{eq:CSLO}\end{eqnarray}
Using this relation, we obtain with the same accuracy as Eq. (\ref{eq:W2})
the following form \begin{eqnarray}
\frac{dW}{d\Omega} & = & \frac{\varepsilon^{2}}{\left(2\pi\right)^{2}}\int d\boldsymbol{\rho}e^{-i\mathbf{q}\cdot\boldsymbol{\rho}}\exp\left[-nL\int d^{2}Q\left(1-e^{i\mathbf{Q}\cdot\boldsymbol{\rho}}\right)\frac{d\sigma}{d\mathbf{Q}}\right]\nonumber \\
 &  & \times\biggl\{1-\frac{n^{2}L}{\varepsilon}\int d\boldsymbol{\rho}_{1}\cos\left(\chi\left(\boldsymbol{\rho}_{1}\right)-\chi\left(\boldsymbol{\rho}_{1}-\boldsymbol{\rho}\right)\right)\boldsymbol{\rho}_{1}\chi\left(\boldsymbol{\rho}_{1}\right)\nonumber \\
 &  & \times\int d\boldsymbol{\rho}_{2}\,\sin\left(\chi\left(\boldsymbol{\rho}_{2}\right)-\chi\left(\boldsymbol{\rho}_{2}-\boldsymbol{\rho}\right)\right)\nabla_{\boldsymbol{\rho}_{2}}\chi\left(\boldsymbol{\rho}_{2}\right)\biggl\}.\label{eq:Wfinal}\end{eqnarray}
The leading term $\frac{dW_{M}}{d\Omega}$ in Eq. (\ref{eq:Wfinal}), corresponding to unity in braces, coincides with
the Moli\`ere's formula. The correction $\frac{dW_{C}}{d\Omega}$ describes the effect of the bulk density of the target
and has not been known so far.

\section{Discussion}

Let us discuss the magnitude and the structure of the correction obtained.
At fixed areal density $nL$ (the number of target atoms per unit
area), the correction behaves as $n$ (or $L^{-1}$), and increases
when $L$ decreases. Estimations show that the relative magnitude
of the correction is the largest when the main contribution to the
integral over $\boldsymbol{\rho}$ in Eq. (\ref{eq:Wfinal}) comes
from the region $\rho\sim a$, where $a$ is the screening radius
of atom, $a\sim a_{B}Z^{-1/3}$, $a_{B}$ is the Bohr radius, $Z$
is the nuclear charge number. This condition is fulfilled when $q\sim nLa$,
where $q$ is the momentum transfer. In this case, the correction
has the relative order\begin{equation}
\delta=\left(\frac{dW_{M}}{d\Omega}\right)^{-1}\frac{dW_{C}}{d\Omega}\sim\frac{Z\alpha na^{3}}{\varepsilon a}R,\quad R=\left(Z\alpha\right)^{2}nLa^{2},\label{eq:relcor}\end{equation}
where $\alpha=1/137$ is the fine-structure constant. Using the estimates\begin{gather*}
\left(\varepsilon a\right)^{-1}\ll\left(ma\right)^{-1}\sim\alpha Z^{1/3}\ll1,\\
Z\alpha na^{3}\lesssim Z\alpha a_{B}^{-3}\left(a_{B}Z^{-1/3}\right)^{3}=\alpha\ll1,\end{gather*}
we obtain\[
\delta\lesssim10^{-3}R\frac{m}{\varepsilon}.\]
The upper bound for $\delta$ grows with $R$. However, when $R\gg1$,
both the leading term and the correction are suppressed by the factor
$\exp\left[-b\, R\right],$ where $b\sim1$ is some numerical constant.
Therefore, in the whole region interesting from the experimental point
of view $R$ is not too big so that $\delta$ is very small.

To conclude, we have calculated the volume density correction to the Moli\`ere's formula and estimated its magnitude.
This correction turns out to be very small for all reasonable values of parameters. Therefore, the Moli\`ere's formula
remains very accurate even for high density of the target.

\section*{Acknowledgements}

The work was supported by grants RFBR 08-02-91969 and DFG GZ 436 RUS
113/769/0-2.

\bibliographystyle{unsrt}

\end{document}